\documentclass[%
longbibliography,
prapplied,
amsmath,amssymb,
reprint,
superscriptaddress,
floatfix
]{revtex4-2}

\usepackage{graphicx}
\usepackage{siunitx}
\usepackage[utf8]{inputenc}
\usepackage{amsmath}
\usepackage{amsfonts}
\usepackage{natbib}
\usepackage{xcolor}
\usepackage{caption}
\usepackage{subcaption}
\usepackage{float}
\usepackage{overpic}
\usepackage{ragged2e}

\graphicspath{{figures/}}

\begin{document}

\title{Radio frequency readout and control of Ge/SiGe hole spin qubits with a global accumulation gate}
\date{\today}

\author{Tien-Ho Chang}
\affiliation{Department of Physics, National Tsing Hua University, Taiwan}
\author{Chi-Wei Lee}
\affiliation{Department of Physics, National Tsing Hua University, Taiwan}
\author{Jian-Chang Zeng}
\affiliation{Department of Physics, National Tsing Hua University, Taiwan}
\author{Chia-Hao Wei}
\affiliation{Department of Physics, National Tsing Hua University, Taiwan}
\author{Ching-Shiang Wang}
\affiliation{Department of Physics, National Tsing Hua University, Taiwan}
\author{Fu-Yuan Gu}
\affiliation{Department of Physics, National Tsing Hua University, Taiwan}
\author{Guan-Yu Yang}
\affiliation{Department of Physics, National Tsing Hua University, Taiwan}
\author{Ruei-Syuan Chiang}
\affiliation{Department of Physics, National Tsing Hua University, Taiwan}
\author{Ho-Chun Wu}
\affiliation{Institute of Electronics Engineering, National Tsing Hua University, Taiwan}
\author{Ming-Hao Lee}
\affiliation{Center for Condensed Matter Sciences, National Taiwan University, Taiwan}
\author{Ming-Wen Chu}
\affiliation{Center for Condensed Matter Sciences, National Taiwan University, Taiwan}
\author{Guang Li Luo}
\affiliation{Taiwan Semiconductor Research Institute, National Institute of Applied Research, Taiwan}
\author{Ta-Chun Cho}
\affiliation{Taiwan Semiconductor Research Institute, National Institute of Applied Research, Taiwan}
\author{Shawn S. H. Hsu}
\affiliation{Institute of Electronics Engineering, National Tsing Hua University, Taiwan}
\affiliation{Department of Electrical Engineering, National Tsing Hua University, Taiwan}
\author{Tzu-Kan Hsiao}
\email{tkhsiao@phys.nthu.edu.tw}
\affiliation{Department of Physics, National Tsing Hua University, Taiwan}

\begin{abstract}

Hole spin qubits in undoped Ge/SiGe quantum well structures have advanced rapidly in performance and scalability. However, stringent multi-layer patterning and overlay requirements of conventional overlapping-gate devices create a bottleneck for academic proof-of-concept experiments involving few-qubit devices. Here we present fabrication and measurements of Ge/SiGe spin qubit devices with a global accumulation gate and single-layer depletion fine gates, which substantially reduce fabrication complexity. With careful design of the gate-2DHG capacitance, we demonstrate RF-based single-shot spin readout and coherent control of two single-spin qubits. We also characterize the spin coherence times and exchange tunability, which are similar to those reported in recent overlapping-gate Ge/SiGe spin qubit devices. By simplifying fabrication without sacrificing performance, our approach offers a more accessible device design for spin-based quantum technology research.

\end{abstract}

\maketitle

Hole spin qubits in undoped Ge/SiGe quantum wells have become a versatile platform for quantum information processing~\cite{Scappucci2021}. Recent advances in material growth and qubit control have facilitated progress in qubit uniformity~\cite{Seidler2025,Stehouwer2025,Tosato2026}, fidelity~\cite{Lawrie2023,Wang2024,Hendrickx2024,Tsoukalas2026}, and scalability~\cite{Borsoi2023,Zhang2025,John2025,Dijkema2026}. Additionally, the strong spin-orbit interaction not only enables all-electrical spin manipulation~\cite{Hendricks2020} but also makes Ge hole spins a promising candidate for spin-photon and spin-phonon hybrid systems~\cite{DePalma2024,Kang2024,Janik2025,Mei2025,Yuan2025}. The conventional overlapping-gate architecture consists of at least four electron-beam (ebeam) defined metal layers (ohmic, screening, plunger, and barrier layer) with nanometer overlay precision and three dielectric thin films in between~\cite{Zajac2016}. Due to complexity, experimental progress in material optimization~\cite{Stehouwer2025,Tosato2026}, cryo-electronics integration~\cite{Bartee2025}, automatic tuning~\cite{Rao2025}, and hybrid systems~\cite{Kang2024} is often limited by the fabrication turnaround and accessibility of spin qubit devices. Therefore, a simple and robust device gate stack with uncompromised functionalities will benefit the research of spin-based quantum technologies.

Early spin qubits based on doped GaAs/AlGaAs heterostructures have the simplest gate stack~\cite{Ciorga2000,Elzerman2004,Petta2005}, which consists of annealed ohmic contacts and subsequent depletion gates. Nevertheless, their qubit coherence is limited by hyperfine noise. In undoped Si or Ge heterostructures, one can make similar 'depletion-like' devices by adding a global accumulation gate (hereafter referred to as global gate) for inducing charge carriers in the active region and then forming quantum dots using depletion gates beneath the global gate. Si/SiGe devices based on the global-gate design have demonstrated coherent spin control and radio-frequency (RF) reflectometry readout~\cite{Noiri2020}. However, Si/SiGe device fabrication typically necessitates a tight gate pitch of 50-100\,nm, ion implantation for ohmic contacts, and micromagnets or ESR lines for spin manipulation.
In contrast, thanks to the small hole effective mass, high-mobility two-dimensional hole gas (2DHG), annealed ohmic contacts, and intrinsic spin-orbit coupling, Ge/SiGe devices based on the global-gate design can be fabricated in a process almost as simple as the one for GaAs devices. Recently, a coherent singlet-triplet qubit has been demonstrated in an undoped global-gated Ge/SiGe device~\cite{Rooney2025}. Nonetheless, the qubit was operated using baseband pulses and measured by averaged sensor currents. In addition, the decoherence time is 600\,ns, which is shorter than that of singlet-triplet qubits defined in overlapping-gate devices~\cite{Zhang2025} and is possibly due to out-of-plane magnetic field component~\cite{Hendrickx2024,Stehouwer2025}. So, it remains unexplored whether or not the global-gate capacitance is compatible with RF readout and microwave(MW)-driven spin control in Ge/SiGe devices~\footnote{Global-gate-free depletion-mode Ge/SiGe devices with unexpected oxide-induced 2DHG have shown RF reflectometry and MW spin control~\cite{Jirovec2021,Saez2025}. However, global-gated devices are more compatible with the majority of state-of-the-art undoped Ge/SiGe heterostructures~\cite{Stehouwer2023,Myronov2025,Costa2026}.}, and if or not the large area of 2DHG under the global gate degrades the coherence of the spin qubits.

\begin{figure*}[t]
    \centering
    \begin{subfigure}[b]{0.32\textwidth}
        \centering
        \begin{overpic}[width=\textwidth]{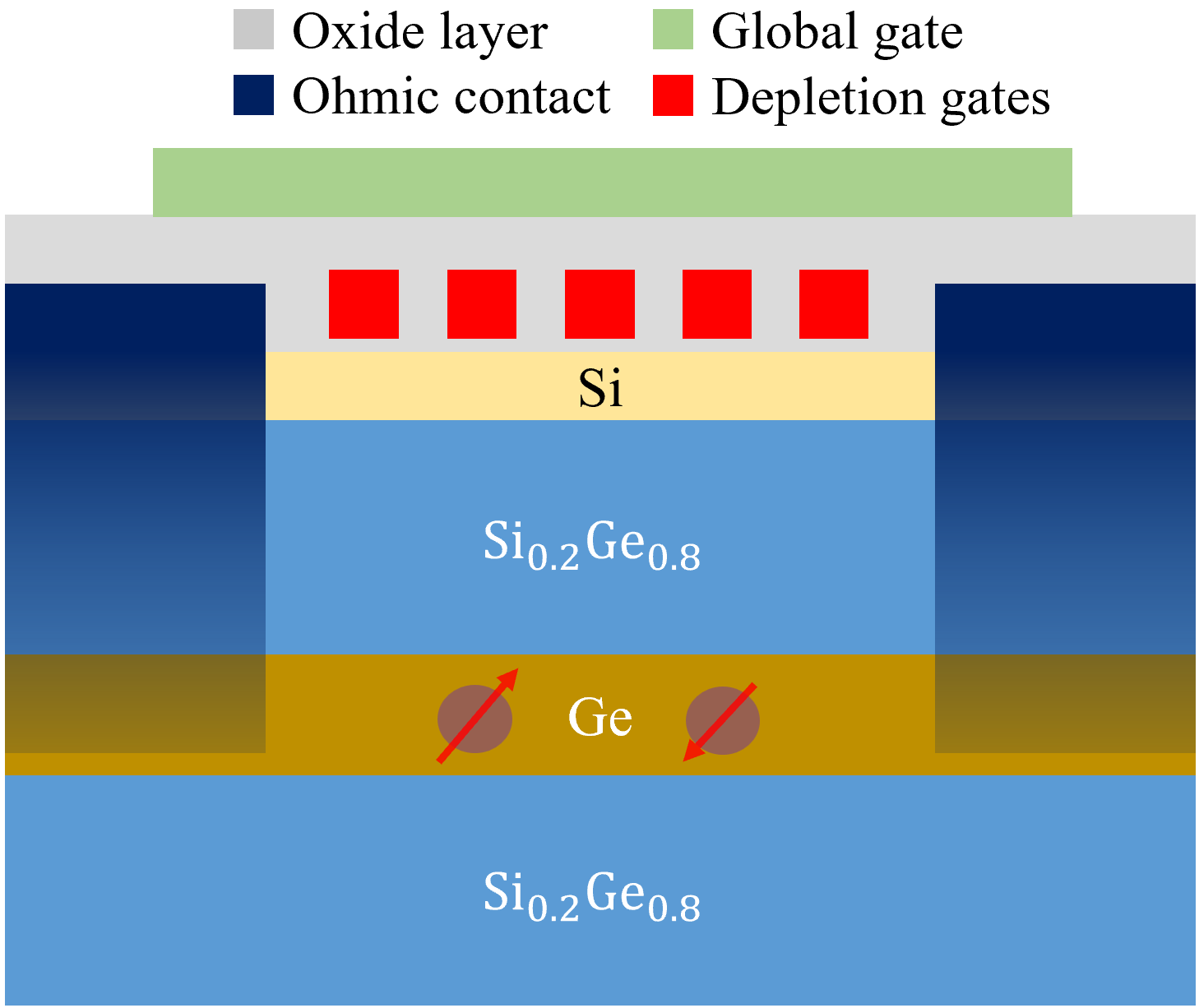}
          \put(3,88){\textbf{(a)}}
        \end{overpic}
    \end{subfigure}
    \hfill
    \begin{subfigure}[b]{0.32\textwidth}
        \centering
        \begin{overpic}[width=\textwidth]{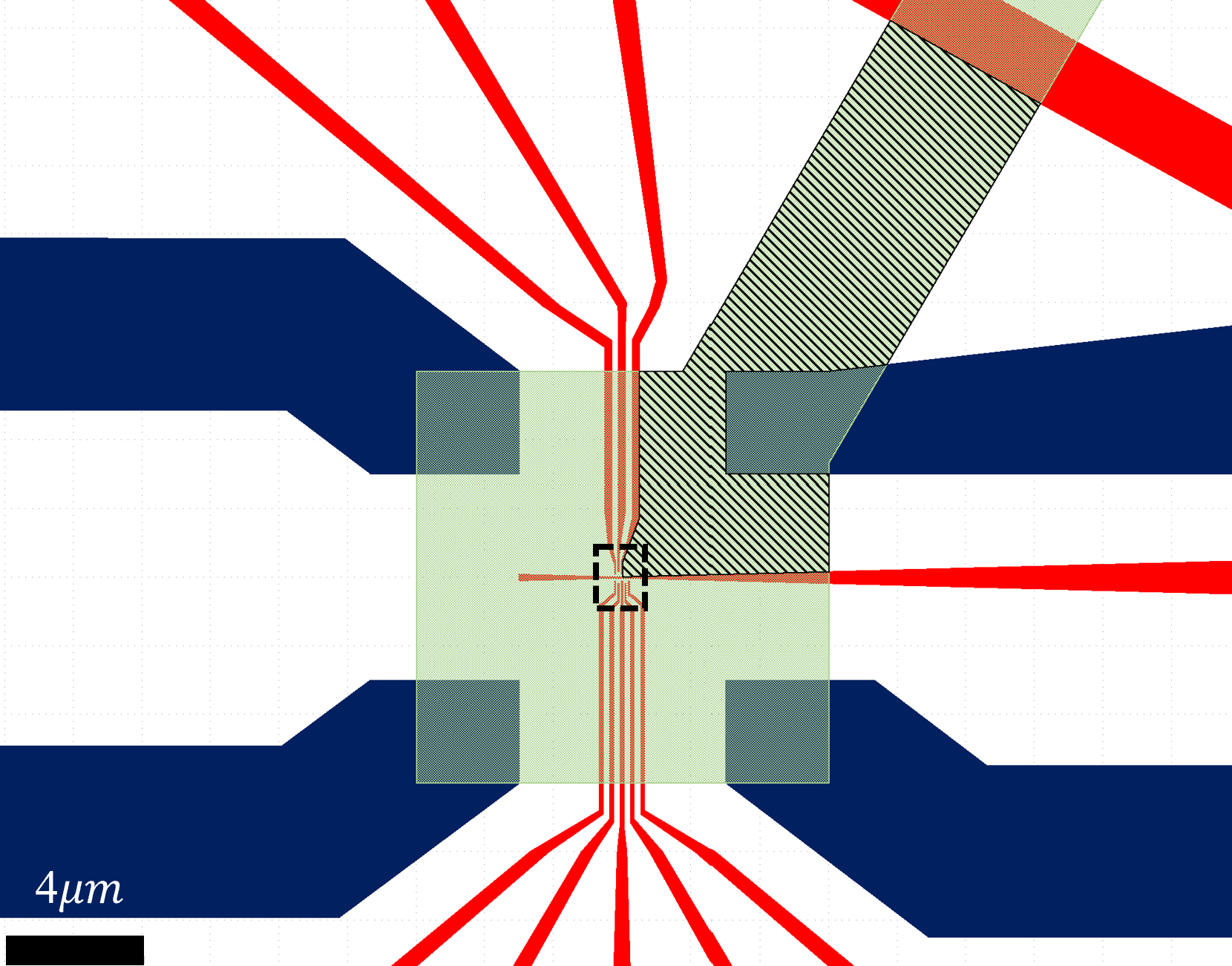}
          \put(3,88){\textbf{(b)}}
        \end{overpic}
    \end{subfigure}
    \hfill
    \begin{subfigure}[b]{0.32\textwidth}
        \centering
        \begin{overpic}[width=\textwidth]{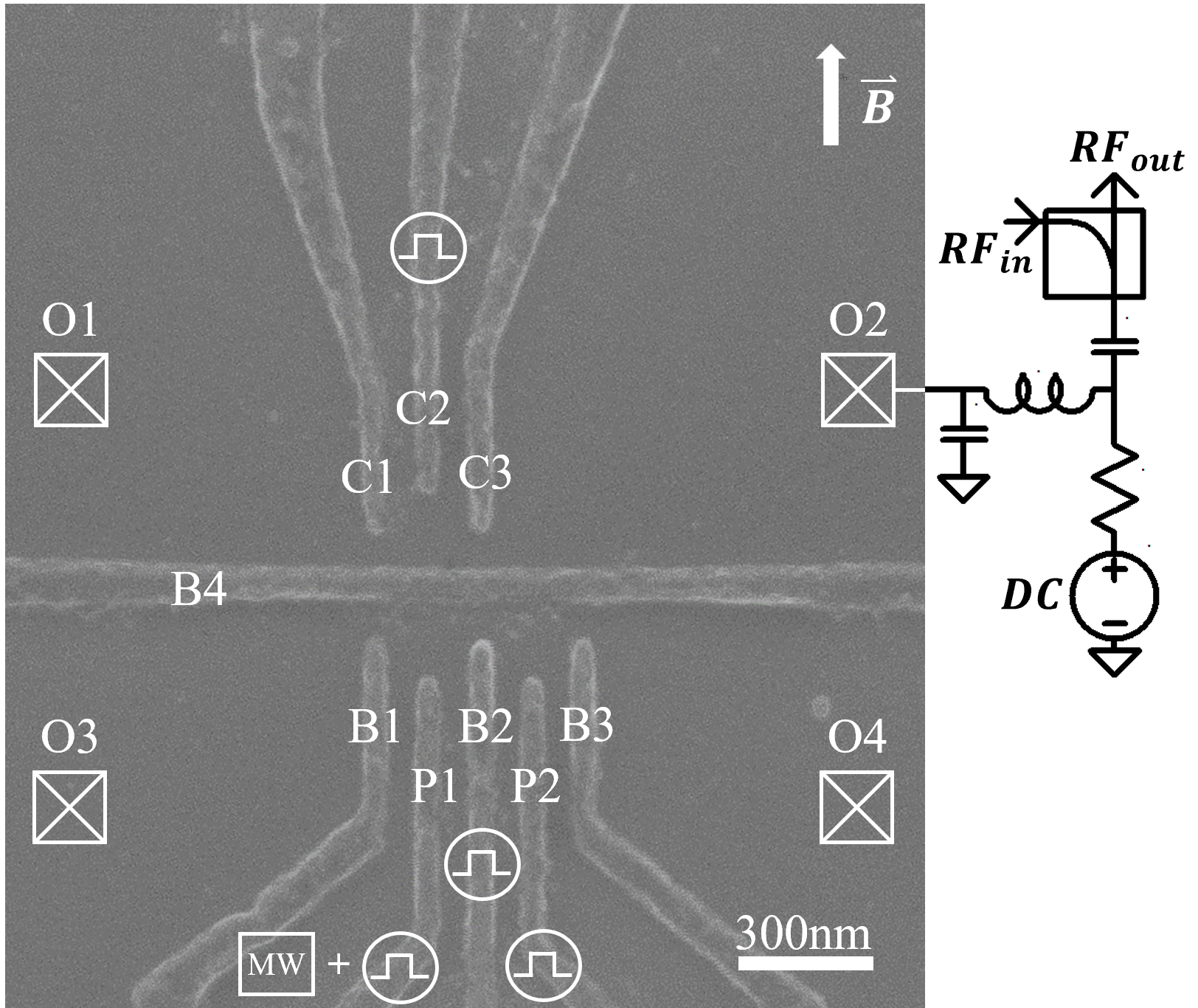}
          \put(3,88){\textbf{(c)}}
        \end{overpic}
    \end{subfigure}
    \caption{
    \justifying
    (a) Schematic cross section of the gate stack and a Ge/SiGe quantum well structure. The gate stack consists of annealed Pt ohmic contacts, depletion Al gates, and a global Al gate. Oxide layers are grown in between metal layers. Hole spins are confined in the quantum well. Colors for each gate layer are shown in the legend. (b) Top view of the gate layout near the active region. Gate layers with different functions are drawn according to the same colors in (a). The shaded area indicates the 2DHG which contributes additional capacitance when performing RF reflectometry via $O_2$ ohmic.
    (c) Scanning electron microscope image showing the ebeam-defined depletion gates in the white rectangular area in (b) before deposition of the global gate. The charge sensor is defined using $C_1-C_3$. The DQD is defined using $P_1$, $P_2$, and $B_1-B_3$. Fast pulses are applied to $C_2$, $P_1$, $P_2$, and $B_2$. MW is applied to $P_1$ for spin control.}
    \label{Device_overveiw}
\end{figure*}

In this work, we demonstrate RF reflectometry single-shot spin readout and MW-driven spin control on Ge/SiGe hole spin qubits with the global-gate design. Specifically, we first show a device design which requires only one ebeam-defined layer with a gate pitch of 100\,nm, two optically defined layers, and a micrometer-scale overlay accuracy between layers. The global gate contributes a gate-2DHG capacitance less than 0.1\,pF. Next, we employ RF reflectometry for charge and spin readout. Finally, we perform MW-driven coherent single spin manipulation, measure exchange coupling, and obtain coherence time and exchange tunability comparable to state-of-the-art overlapping-gate Ge/SiGe devices.

Our Ge hole spin qubit devices are fabricated on a chemical-vapor-deposition-grown reverse-graded Ge/SiGe quantum well structure with a hole mobility of $1.0\times10^5$\,cm$^2$V$^{-1}$s$^{-1}$ at a density of $1.5\times10^{11}$\,cm$^{-2}$ and a percolation density of $3.1(1)\times10^{10}$\,cm$^{-2}$ (see supplementary material for details). As shown in Fig.~\ref{Device_overveiw}(a), from bottom to top the gate stack includes annealed 30\,nm platinum ohmic contacts, 30\,nm aluminum depletion gates, and 70\,nm aluminum global gate. Between each metal layer, aluminum oxide is grown using atomic layer deposition (with thickness 50\,nm between the global gate and the depletion gates, and 5\,nm between the depletion gates and the substrate). The depletion gates with a minimum pitch of 100\,nm, as shown in Fig.~\ref{Device_overveiw}(c), are patterned using ebeam lithography while the micrometer-scale ohmic contacts and the 12$\times$12\,\si{\micro\m}$^2$ global gate are defined using photolithography. As shown in Fig.~\ref{Device_overveiw}(b), functional device can be obtained as long as the depletion gates are located in the middle of the ohmics and also the global gate covers the ohmics and depletion gates. Therefore, our design relaxes the fabrication requirements to only one layer of ebeam gates with a 100\,nm pitch and an overlay tolerance of several micrometers.

When 2DHG is accumulated in the quantum well by the global gate, the parallel-plate capacitor between the global gate and the 2DHG gives additional capacitance to the RF reflectometry circuit, which typically has a self capacitance in the order of 1\,pF contributed from the printed circuit board (PCB), bond wires, and ohmic contacts.
To make our qubit device compatible with RF readout, we design the global gate to have a relatively small area in the active region while photolithography patterning and alignment are still feasible. When sending RF signal from the $O_2$ ohmic contact, we can estimate a parallel-plate capacitance of 69\,fF from the global gate to the shaded area of 2DHG~\cite{Noiri2020} in Fig.~\ref{Device_overveiw}(b), assuming $\epsilon_\mathrm{Al_{2}O_{3}}=10$, $\epsilon_\mathrm{Si}=11.7$, and $\epsilon_\mathrm{Si_{0.2}Ge_{0.8}}=15.3$. Hence, the additional capacitance from the global gate would only have a minor effect on the impedance matching condition.
Device measurements are performed using a custom-made sample PCB loaded into a dilution refrigerator at a base temperature of 10\,mK. (see supplementary material for setup details). Qubit measurements are carried out under an in-plane magnetic field of 50\,mT.

\begin{figure}[t!]
    \centering
    \begin{subfigure}[b]{0.225\textwidth}
        \centering
        \begin{overpic}[width=\textwidth]{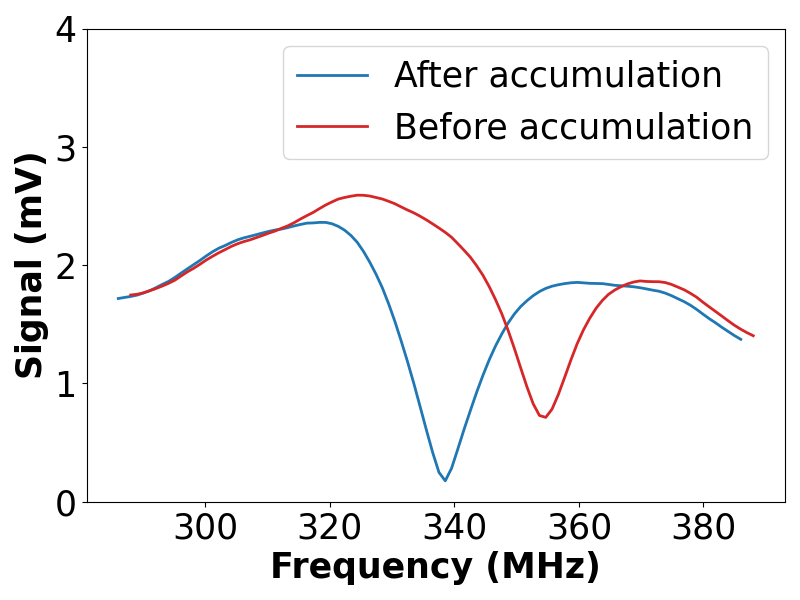}
          \put(3,81){\textbf{(a)}}
        \end{overpic}
        \label{global gate}
    \end{subfigure}
    \hfill
    \begin{subfigure}[b]{0.245\textwidth}
        \centering
        \begin{overpic}[width=\textwidth]{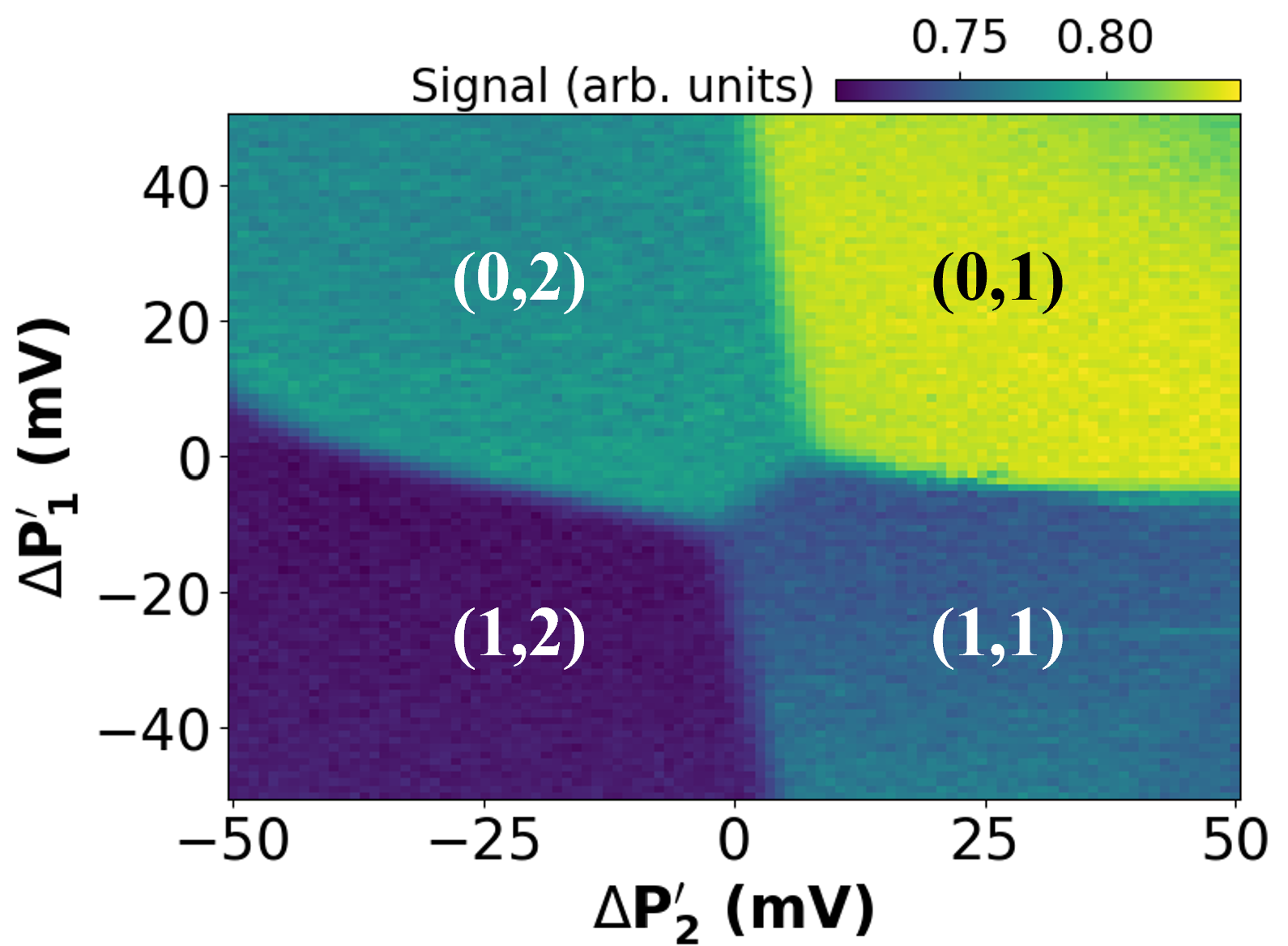}
            \put(3,74){\textbf{(b)}}
        \end{overpic}
        \label{charge stability diagram}
    \end{subfigure}
    \caption{
    \justifying
    (a) resonance frequency of the LC tank circuit before and after the 2DHG accumulation. The frequency shift corresponds to a gate-2DHG capacitance of $73$\,fF  (b) Charge stability diagram near the few-hole regime measured using RF reflectometry.}
    \label{RF_readout}
\end{figure}

To characterize the global-gate capacitance, we mount a surface-mount wirewound ceramic core inductor of 300\,nH on the sample PCB for the LC readout tank. The readout line is connected to the top-right ohmic ($O_2$) of device 1. As shown in Fig.~\ref{RF_readout}(a), when device 1 is grounded the resonance frequency is 354.7\,MHz, indicating a self capacitance of 746\,fF before 2DHG accumulation. The resonance frequency after 2DHG accumulation shifts to 338.5\,MHz, corresponding to a capacitance of 819\,fF. Therefore, the added capacitance of the global gate is 73\,fF, which is consistent with the estimated value.

To demonstrate RF readout and MW control, we measure device 2, which is nominally identical to device 1, using RF reflectometry with a TiN spiral chip inductor. A charge sensor and a double quantum dot (DQD) array are formed by applying DC voltages to the depletion gates labeled in Fig.~\ref{Device_overveiw}(c), in which $C_2$, $P_1$ and $P_2$ primarily control the electrochemical potentials of the sensor, quantum dot 1 and quantum dot 2, respectively, and $B_2$ mainly modulates the tunnel coupling of the DQD. To suppress the capacitive crosstalk between gate electrodes, We define virtual gate voltages $C'_2$, $P'_1$, $P'_2$, and $B'_2$, which are linear combinations of the physical gate voltages~\cite{Hensgens2017}.
Figure~\ref{RF_readout}(b) shows the video-mode charge stability diagram measured by the reflected RF signal from the charge sensor as a function of fast voltage pulses $\Delta P'_1$ and $\Delta P'_2$ relative to DC voltages~\cite{Stehlik2015,Baart2016}. The DQD charge occupation is near the (1,1)-(0,2) transition, where the following spin qubit experiments are performed.

\begin{figure}[b]
    \centering
    \begin{subfigure}[b]{0.23\textwidth}
        \begin{overpic}[width=\textwidth]{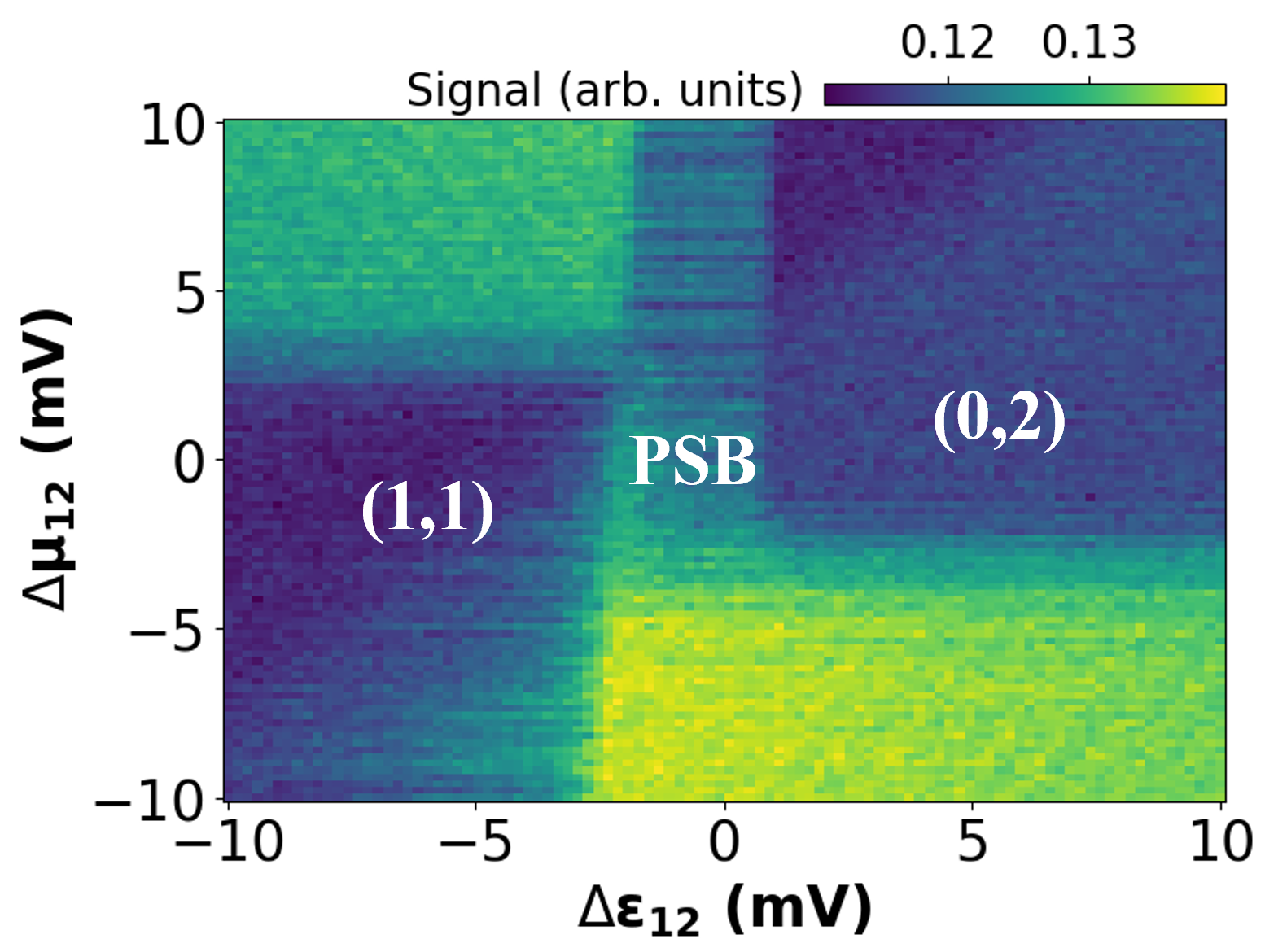}
          \put(3,76){\textbf{(a)}}
        \end{overpic}
        \label{PSB1}
    \end{subfigure}
    \centering
    \begin{subfigure}[b]{0.23\textwidth}
        \centering
        \begin{overpic}[width=\textwidth]{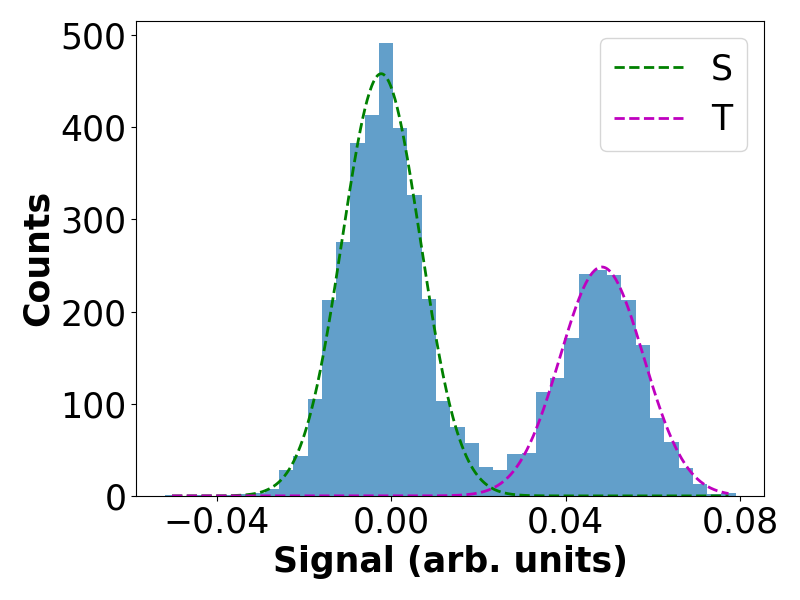}
          \put(3,78){\textbf{(b)}}
        \end{overpic}
        \label{PSB2}
    \end{subfigure}
    \hfill
    \caption{
    \justifying
    (a) PSB readout region on the charge stability diagram near the (1,1)-(0,2) transition. (b) PSB readout histogram showing singlet and triplet distributions with a SNR of 4.6 using an integration time of 10\,\si{\micro\s}.}
    \label{PSB}
\end{figure}

Spin states are measured using Pauli-spin blockade (PSB)~\cite{Barthel2009}. Figure~\ref{PSB}(a) shows the sensor signal (with an integration time of 2\,us per pixel) as a function of detuning $\Delta\epsilon_{12} = \Delta P'_1-\Delta P'_2$ and total energy $\Delta\mu_{12} = \Delta P'_1+\Delta P'_2$ near the (1,1)-(0,2) transition. In Fig.~\ref{PSB}(a) only the interdot transition and a region corresponding to the PSB readout window are clearly visible, since we close the dot-reservoir coupling for better PSB signals. In Fig~\ref{PSB}(b) we perform RF-based single-shot spin readout by randomly loading two spins in the (1,1) charge state and then measuring the spin signal in the PSB readout window. The signlet-triplet double peaks are observed in the readout histogram with a signal-to-noise ratio (SNR) of 4.6 with an integration time of 10\,\si{\micro\s}.

\begin{figure*}[t]
    \centering
    \begin{subfigure}[b]{0.31\textwidth}
        \centering
        \begin{overpic}[width=\textwidth]{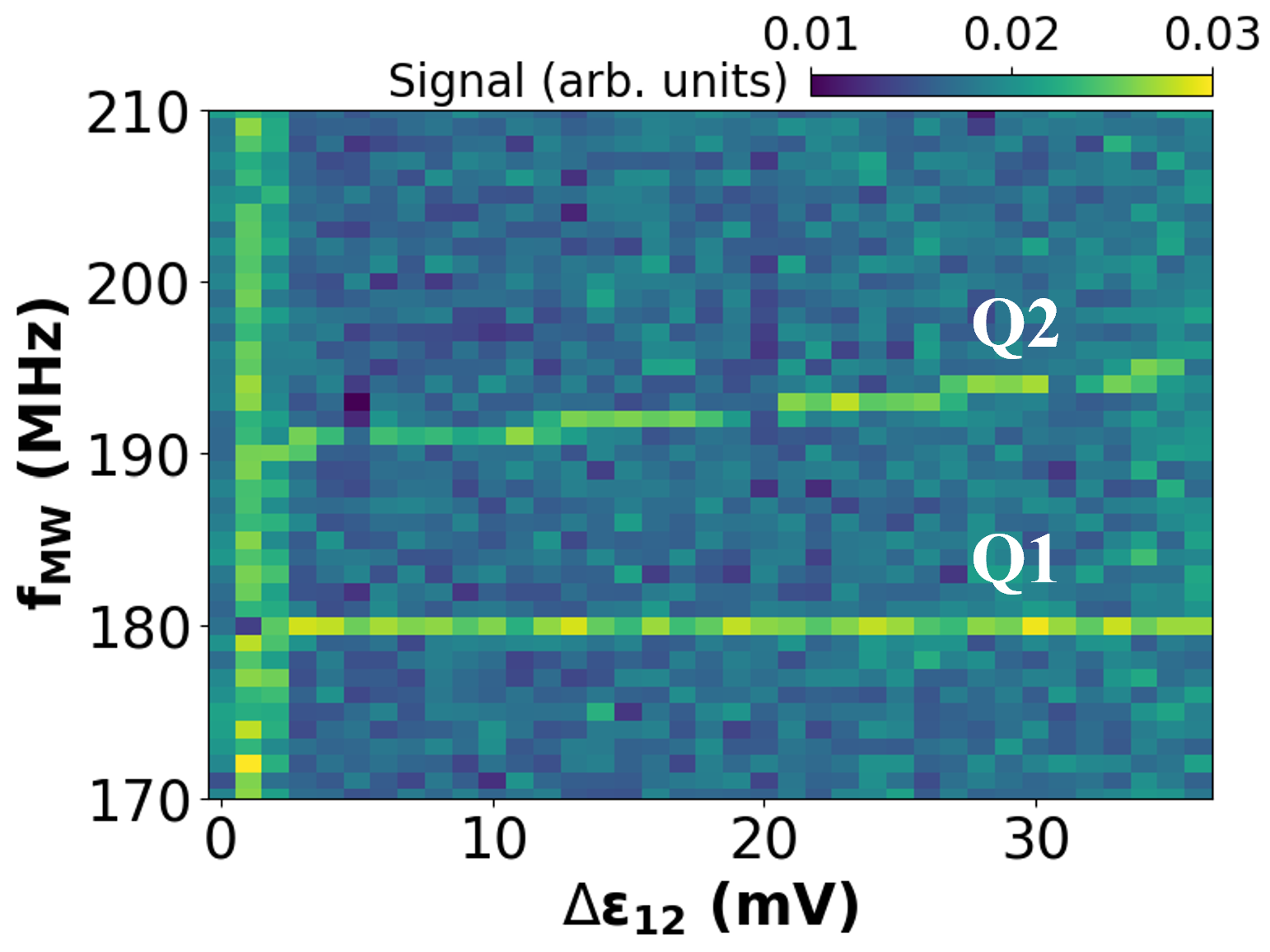}
          \put(3,80){\textbf{(a)}}
        \end{overpic}
        \label{detuning and microwave frequency}
    \end{subfigure}\hfill
    \begin{subfigure}[b]{0.31\textwidth}
        \centering
        \begin{overpic}[width=\textwidth]{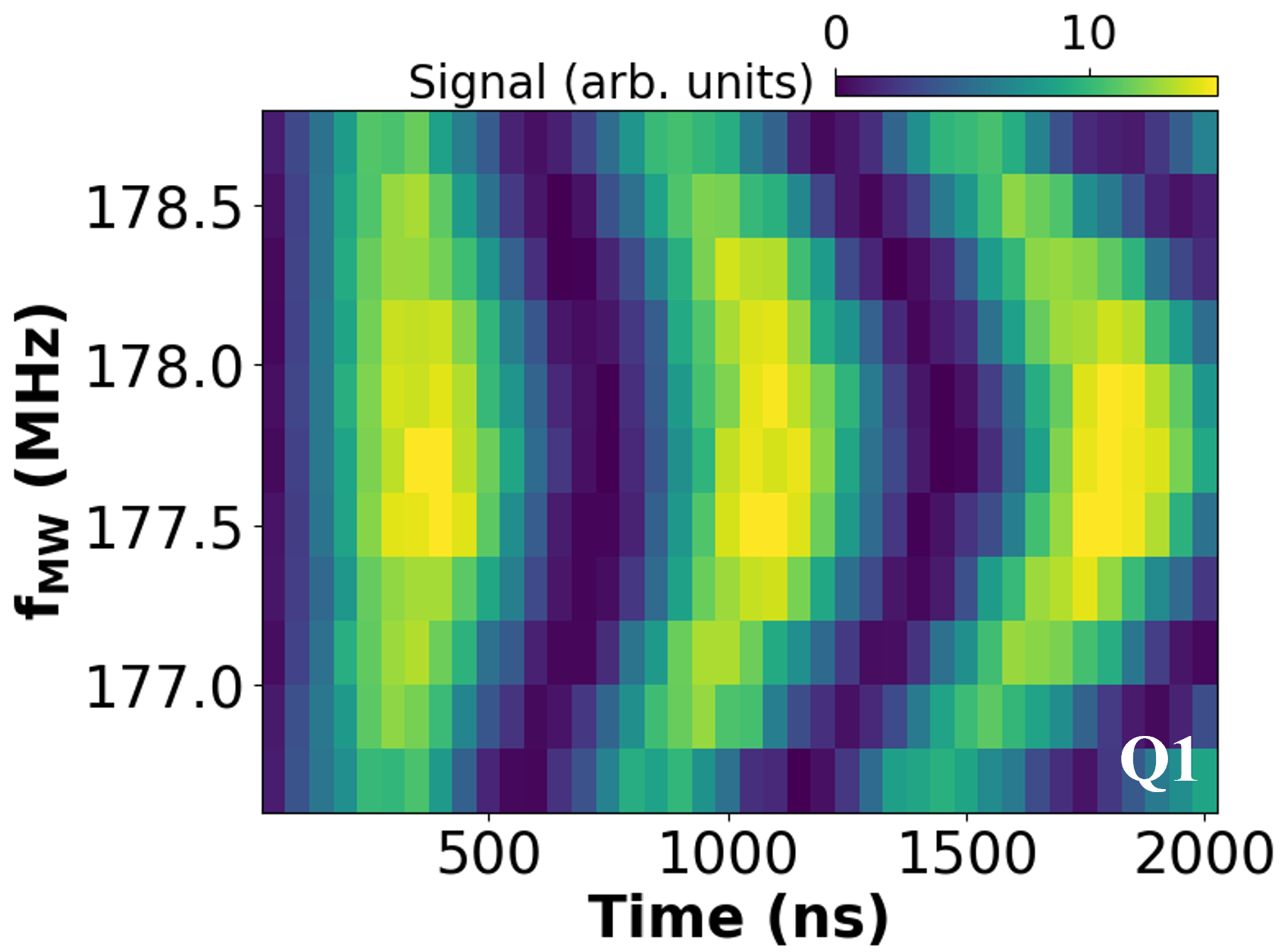}
          \put(3,80){\textbf{(b)}}
        \end{overpic}
        \label{Q1 Rabi large}
    \end{subfigure}\hfill
    \begin{subfigure}[b]{0.31\textwidth}
        \centering
        \begin{overpic}[width=\textwidth]{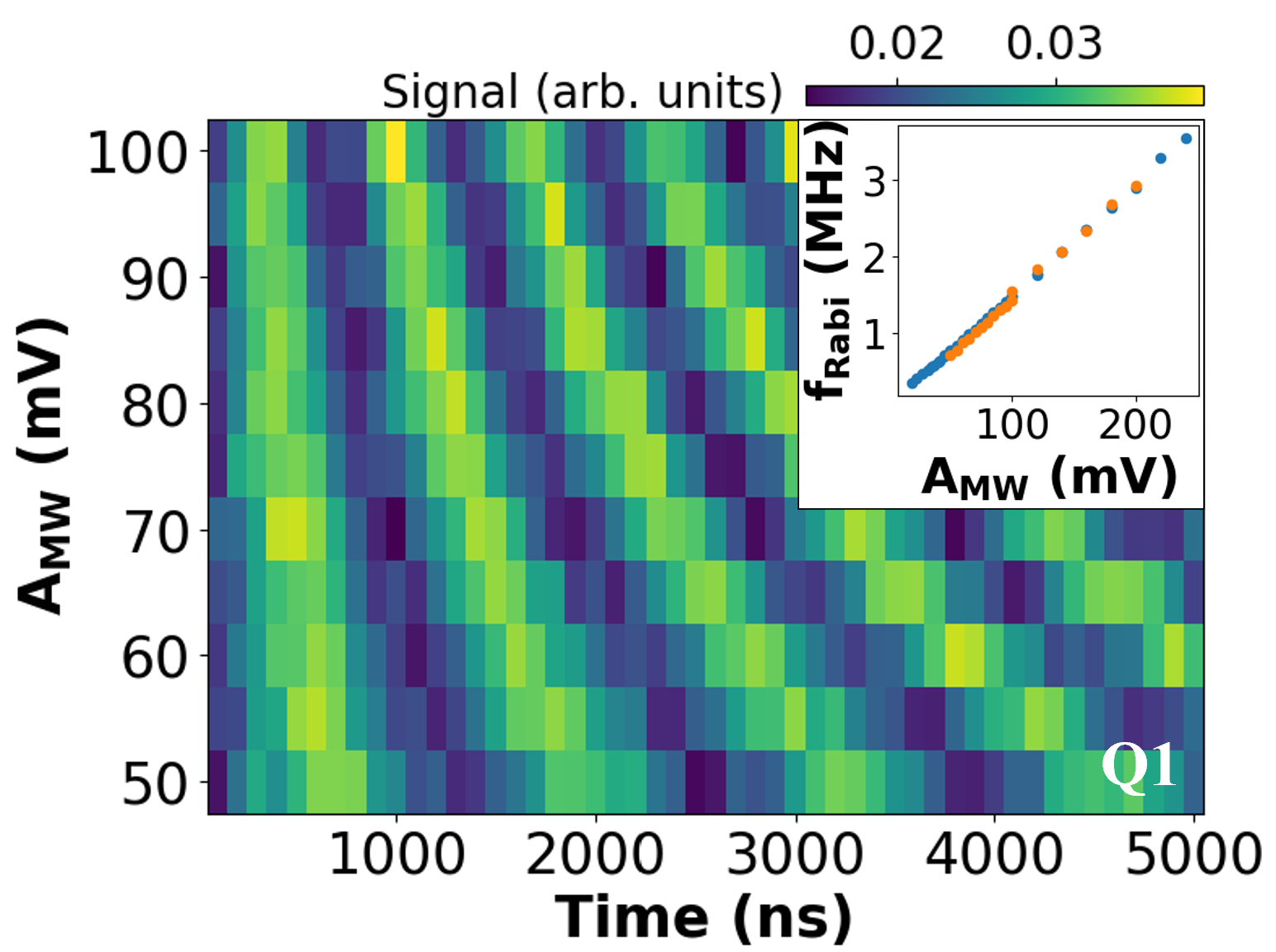}
          \put(3,80){\textbf{(c)}}
        \end{overpic}
        \label{Rabi power}
    \end{subfigure}
    
    \vspace{0.8em} 

    \begin{subfigure}[b]{0.31\textwidth}
        \centering
        \begin{overpic}[width=\textwidth]{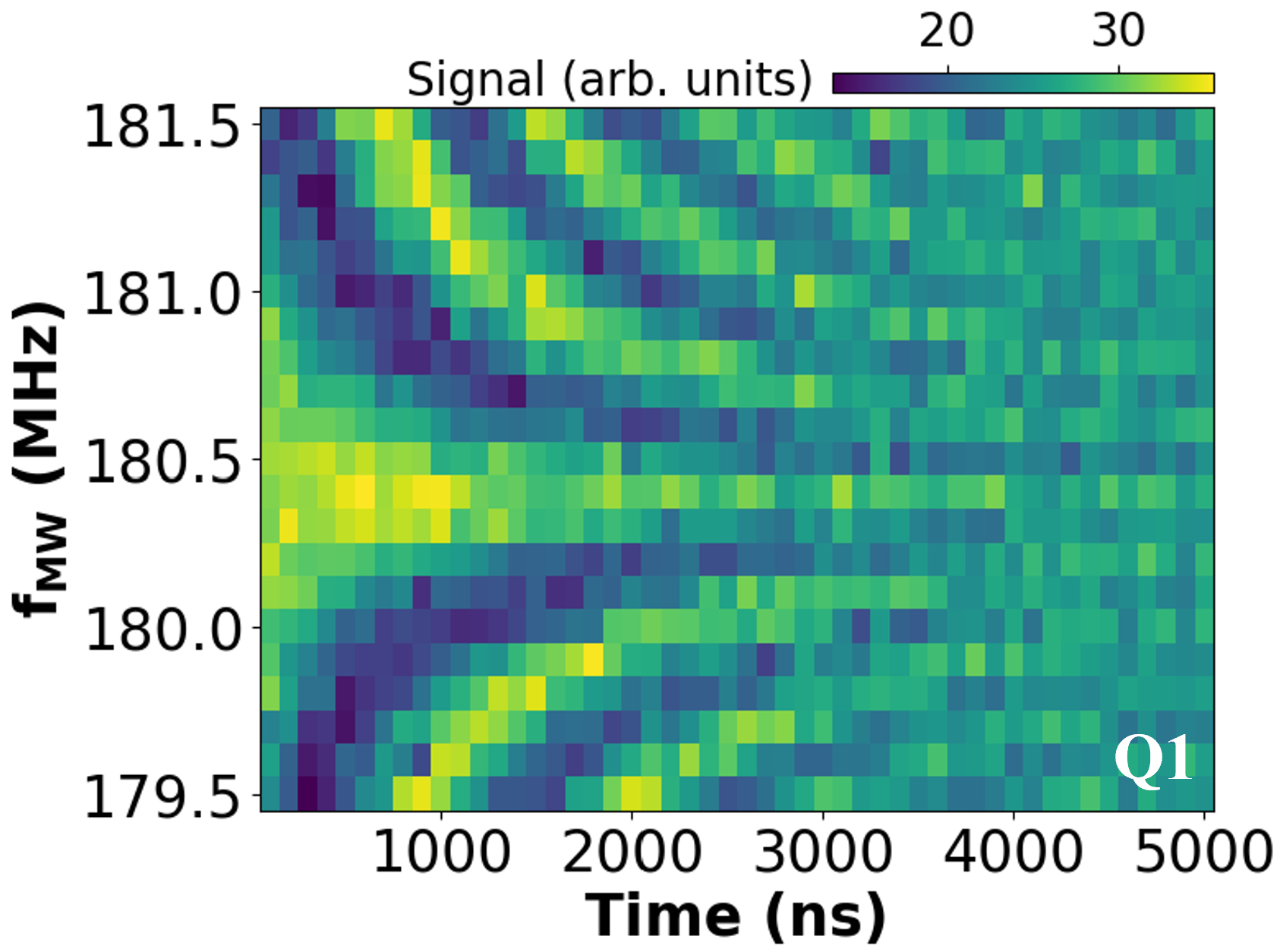}
          \put(3,80){\textbf{(d)}}
        \end{overpic}
        \label{Q1 ramsey}
    \end{subfigure}\hfill
    \begin{subfigure}[b]{0.31\textwidth}
        \centering
        \begin{overpic}[width=\textwidth]{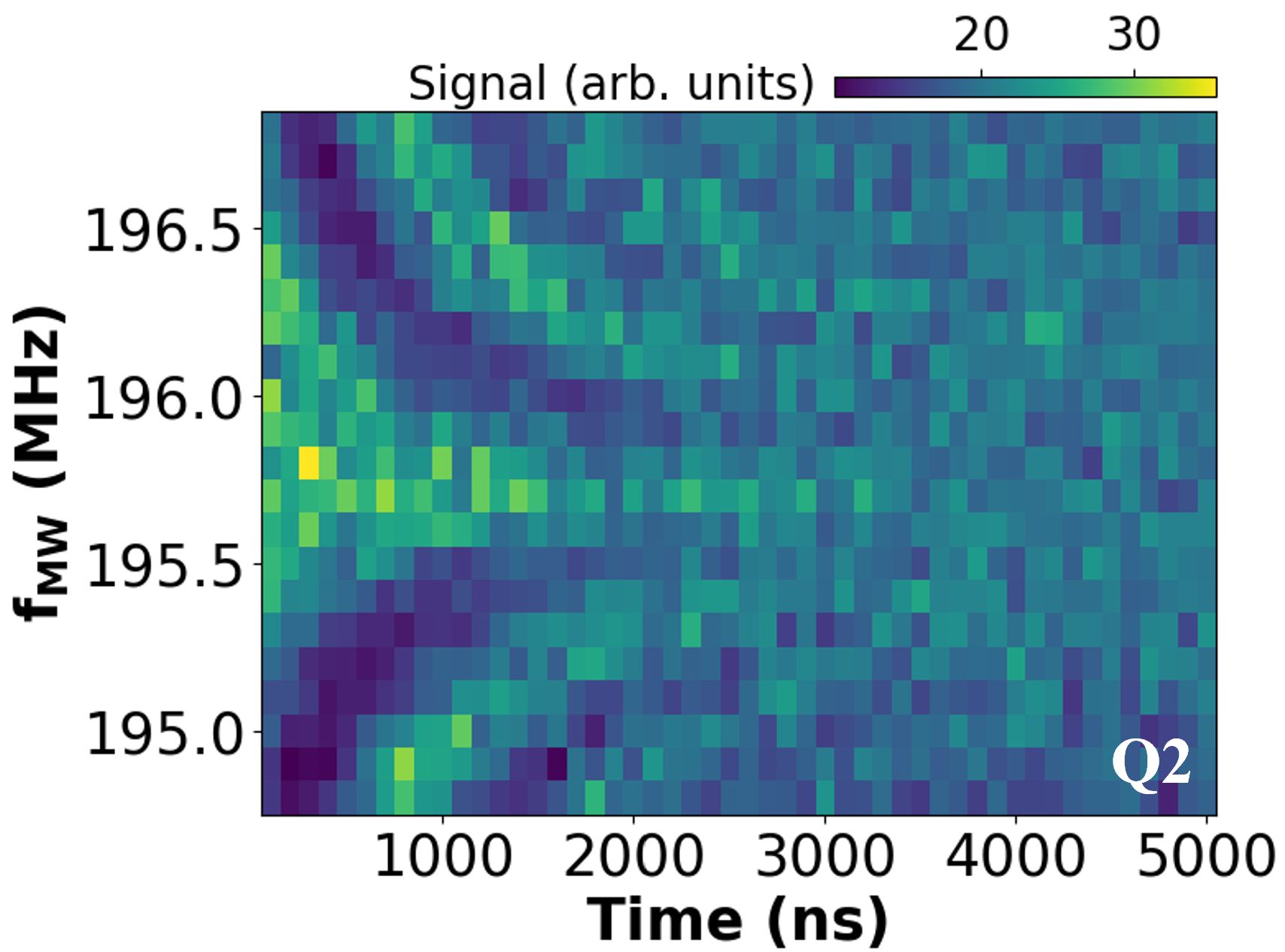}
          \put(3,80){\textbf{(e)}}
        \end{overpic}
        \label{Q2 ramsey}
    \end{subfigure}\hfill
    \begin{subfigure}[b]{0.31\textwidth}
        \centering
        \begin{overpic}[width=\textwidth]{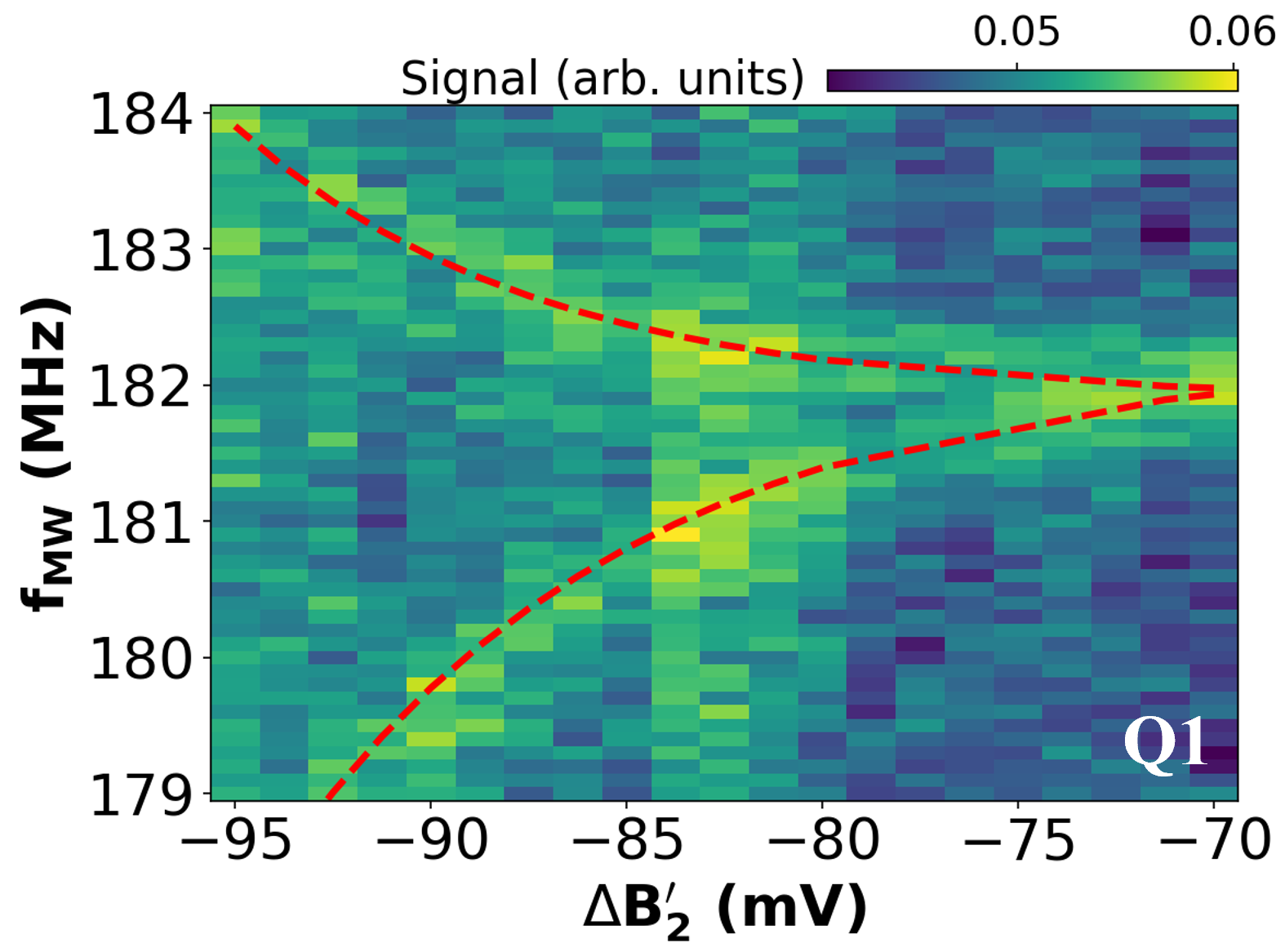}
          \put(3,80){\textbf{(f)}}
        \end{overpic}
        \label{exchange splitting}
    \end{subfigure}

    \caption{
    \justifying
    (a) MW-excited spin energy spectroscopy as function of $\epsilon_{12} = P'_1-P'_2$ and $f_\mathrm{MW}$. The horizontal lines near $f_\mathrm{MW}=180$\,MHz and 190\,MHz correspond to the resonance frequencies of $Q_1$ and $Q_2$, respectively. The vertical line near zero detuning is the (1,1)-(0,2) transition. (b) MW-driven Rabi oscillation of $Q_1$ near the resonance frequency. (c) Rabi oscillation of $Q_1$ as a function of $A_\mathrm{MW}$. Inset shows the $A_\mathrm{MW}$ dependence of $f_\mathrm{Rabi}$ of $Q_1$ (blue) and $Q_2$ (orange). (d) Ramsey chevron of $Q_1$. (e) Ramsey chevron of $Q_2$. (f) Exchange energy splitting of $Q_1$ as a function of $\Delta B'_2$. The dashed lines are fitting of the exchange splitting to an exponential function, which yields an exchange tunability of $17.3(3)$\,mV/dec.}
    \label{qubit_characterization}
\end{figure*}

To demonstrate coherent MW spin control, we first measure the spin energy spectrum as a function of $\epsilon_{12}$ and the frequency of the MW driving tone $f_\mathrm{MW}$ applied on $P_1$. As shown in Fig.~\ref{qubit_characterization}(a), two horizontal lines are observed, indicating the qubit resonance frequencies of the two spins. Next, we apply MW drive around 180\,MHz as a function of MW driving time and get the Rabi oscillation of qubit 1 ($Q_1$) in Fig.~\ref{qubit_characterization}(b). The Rabi oscillation of $Q_1$ as a function of MW driving amplitude $A_\mathrm{MW}$ is shown in Fig.~\ref{qubit_characterization}(c). Both $Q_1$ and $Q_2$ have a linear dependence of Rabi frequency $f_\mathrm{Rabi}$ on $A_\mathrm{MW}$ (Fig.~\ref{qubit_characterization}(c) inset). To characterize the qubit coherence, we perform Ramsey experiments with a measurement time around 5 minutes. The Ramsey chevrons of $Q_1$ and $Q_2$ are shown in Fig.~\ref{qubit_characterization}(d) and (e), respectively, from which we extract $T^*_2 = 3.15(8)$\,\si{\micro\s} for $Q_1$ and $T^*_2 = 1.94(8)$\,\si{\micro\s} for $Q_2$. The reduced $T^*_2$ of $Q_2$ may be attributed to the detuning-dependent qubit frequency as can be seen in Fig.~\ref{qubit_characterization}(a). These values are comparable with overlapping-gate Ge/SiGe hole spin qubits under similar magnetic field strengths~\cite{Stehouwer2025,John2025,Dijkema2026}.

Exchange coupling $J$ is a typical mechanism for implementing two-qubit operations in spin qubit processors. Here we show the control of exchange coupling in our device. We first prepare a superposition state of $Q_2$ then measure the energy spectrum of $Q_1$ as a function of the interdot barrier $B'_2$. In the regime where Zeeman energy difference $\Delta E_z \gg J$, the resonance frequency of $Q_1$ becomes state-dependent such that $f_1^{Q2\downarrow}=f_1^0 - J/2$ and $f_1^{Q2\uparrow}=f_1^0 + J/2$, where $f_1^0$, $f_1^{Q2\downarrow}$ and $f_1^{Q2\uparrow}$ refer to the $Q_1$ frequencies when $J$ is 0, when $Q_2$ is down and when $Q_2$ is up, respectively~\cite{Watson2018,Zajac2018}. In Fig.~\ref{qubit_characterization}(f), the resonance frequency of $Q_1$ splits into two branches as $B'_2$ is lowered. The energy splitting directly indicates the value of $J$. We then extract the exchange tunability of 17.3(3)\,mV/dec, which is similar to the values from overlapping-gate devices~\cite{Dijkema2026,Wang2024,Jirovec2025}. In addition, we present a $S$-$T_-$ qubit with a coherence time of $T^*_2 = 1.0(1)$\,\si{\micro\s} and charge noise characterization yielding a power spectrum density $S=1.49(8)$\,\si{\micro\eV}/$\sqrt{\mathrm{Hz}}$ at 1\,Hz using nominally identical devices in the supplemental material.


In summary, we have shown a Ge/SiGe hole spin qubit device with a global-gate design. The fabrication process are significantly simplified compared to the overlapping-gate design while the device functionality is not compromised. The small global-gate capacitance of 73\,fF enables single-shot spin readout based on RF reflectometry. From MW-driven coherent spin experiments we obtain $T^*_2$ of 3.1 (1.9)\,\si{\micro\s} for $Q_1$ ($Q_2$) and exchange tunability of $17.3$\,mV/dec, which are consistent with the values reported in state-of-the-art overlapping-gate devices.

Looking ahead, a crucial next step is to analyze gate fidelities and understand error sources of the global-gated devices using benchmarking techniques~\cite{Proctor2025}. Finally, we discuss their application in semiconductor quantum technologies. Although the global-gated design has limited scalability into large 2D arrays, it allows for simple fabrication of linear few-qubit arrays, which provide a more accessible platform for fast-feedback material characterization, training auto-tuning algorithms, and cryo-electronics test. Moreover, few-qubit arrays with this design may work as reliable unit cells for building hybrid systems in Ge/SiGe devices.


\section*{Acknowledgments}
We acknowledge useful discussions with Shangjr Gwo, Jiun-Yun Li, and members of the Hsiao group. We thank the support from the National Science and Technology Council (NSTC) grants (112-2112-M-007-054-MY3) and (114-2119-M-007-008). We also acknowledge Yushan Fellow Program (MOE-111-YSFMS-0002-001-P1) and the Center for Quantum Science and Technology (CQST) by the Ministry of Education (MOE), Taiwan for the financial support.

%

\end{document}



\title{\textbf{Radio frequency readout and control of Ge/SiGe hole spin qubits with a global accumulation gate supplementary material}}
\date{\today}

\author{Tien-Ho Chang \and
        Chi-Wei Lee \and
        Jian-Chang Zeng \and
        Chia-Hao Wei \and
        Ching-Shiang Wang \and
        Fu-Yuan Gu \and
        Guan-Yu Yang \and
        Ruei-Syuan Chiang \and
        Ho-Chun Wu \and
        Ming-Hao Lee \and
        Ming-Wen Chu \and
        Guang Li Luo \and
        Ta-Chun Cho \and
        Shawn S. H. Hsu \and
        Tzu-Kan Hsiao}
 
\maketitle
\tableofcontents
\bigskip\hrule\bigskip

\newpage
\section{Measurement setup}
\label{measurement setup}
\begin{figure}[htbp]
    \centering
    \includegraphics[width=0.9\textwidth]{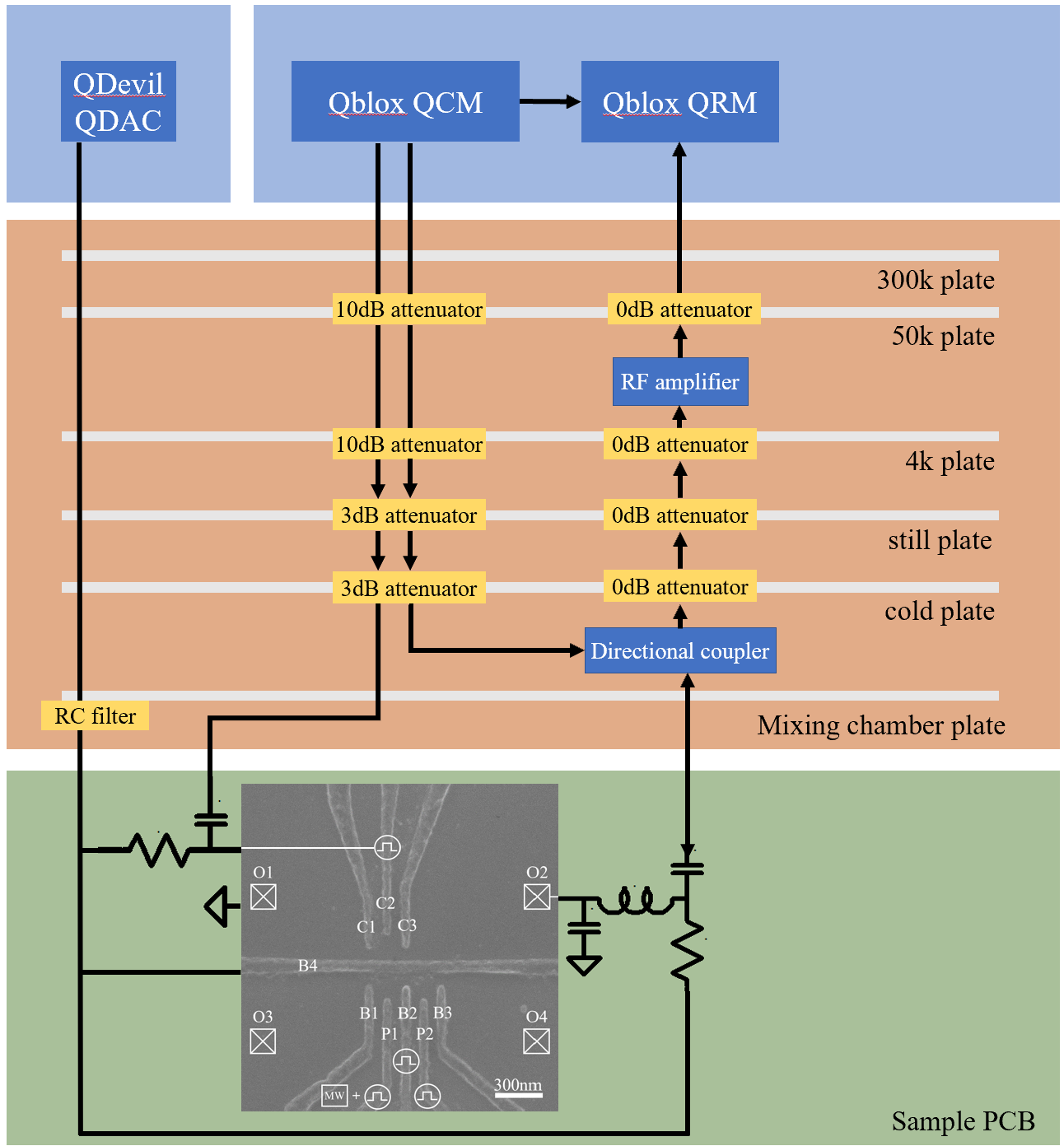}
    \caption{\textbf{Signal circuit diagram}
    showing routing of RF and DC signals for the experiment.}
\end{figure}

Our Device is wire bonded onto a custom made PCB that was designed in house that allows for 48 DC channels and 8 RF input channels and 1 RF output channel. The PCB is loaded into an Oxford Proteox MX450 dilution refrigerator and cooled to 10\,mK. RF signal generation and readout demodulating is handled by the Qblox cluster series qubit control and readout unit. DC signals are provided by two QDACs. The amplifier and directional coupler used in the readout circuit is the CITLF3 cryogenic SiGe low noise amplifier from Cosmic Microwave Technology, Inc. and ZFDC-20-33-S+ from Mini-Circuits respectively. 

\newpage
\section{Material characterization}
\label{mobility}
\begin{figure}[h]
    \centering
    \begin{subfigure}{0.44\textwidth}
        \centering
        \includegraphics[width=\linewidth]{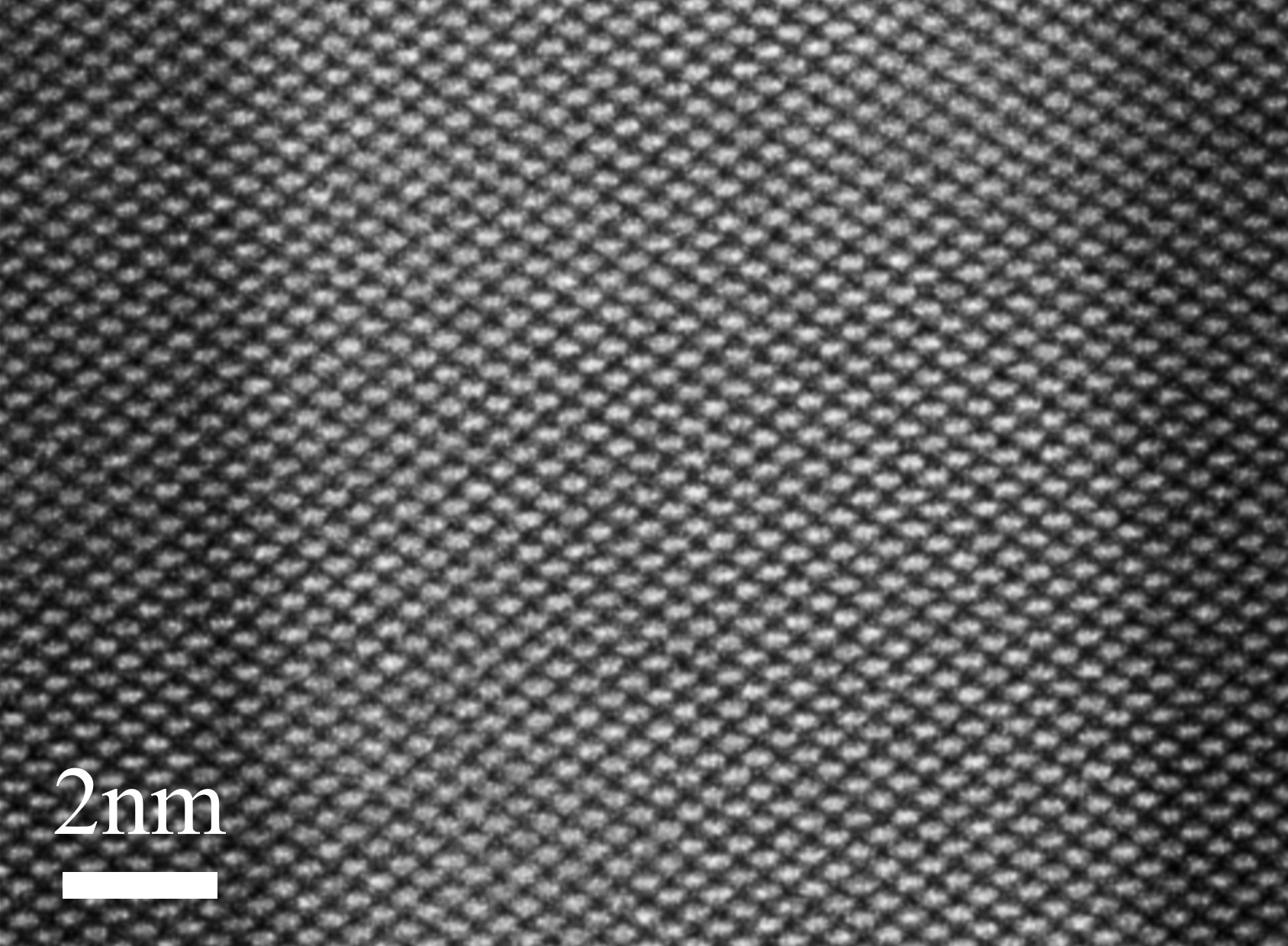}
        \caption{}
        \label{TEM}
    \end{subfigure}
    \hfill
    \begin{subfigure}{0.45\textwidth}
        \centering
        \includegraphics[width=\linewidth]{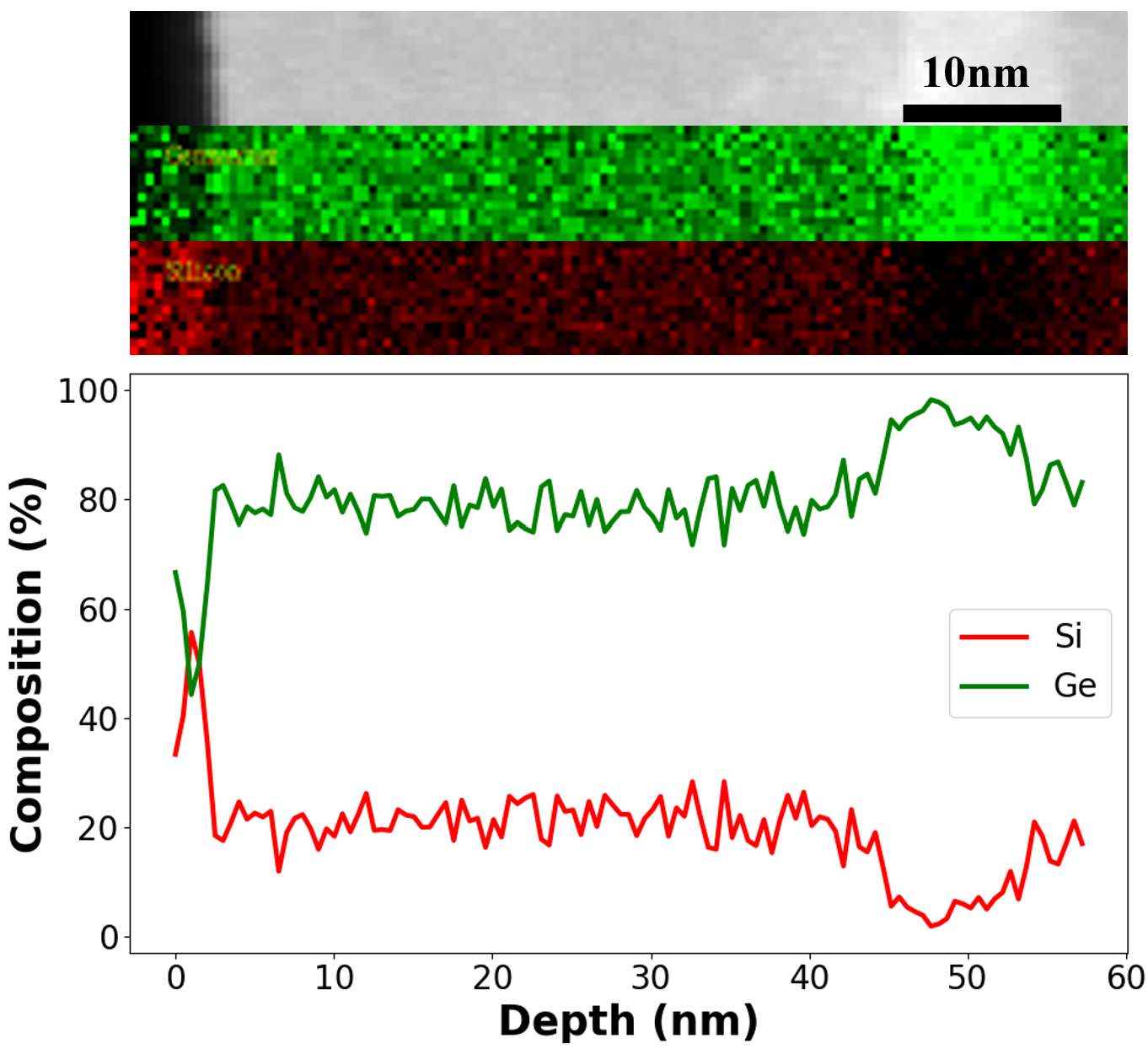}
        \caption{}
        \label{fig:sub2}
    \end{subfigure}

    \vspace{1em}

    \begin{subfigure}{0.44\textwidth}
        \centering
        \includegraphics[width=\linewidth]{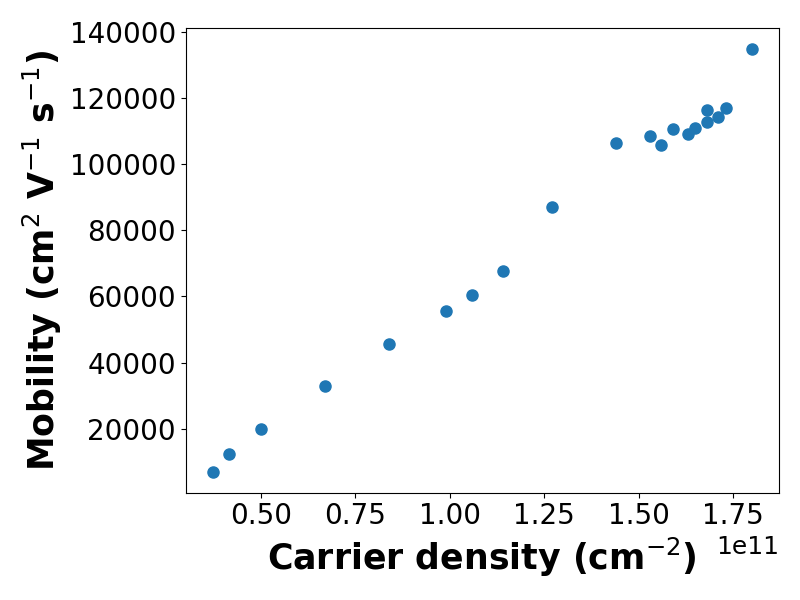}
        \caption{}
        \label{mobility}
    \end{subfigure}
    \hfill
    \begin{subfigure}{0.45\textwidth}
        \centering
        \includegraphics[width=\linewidth]{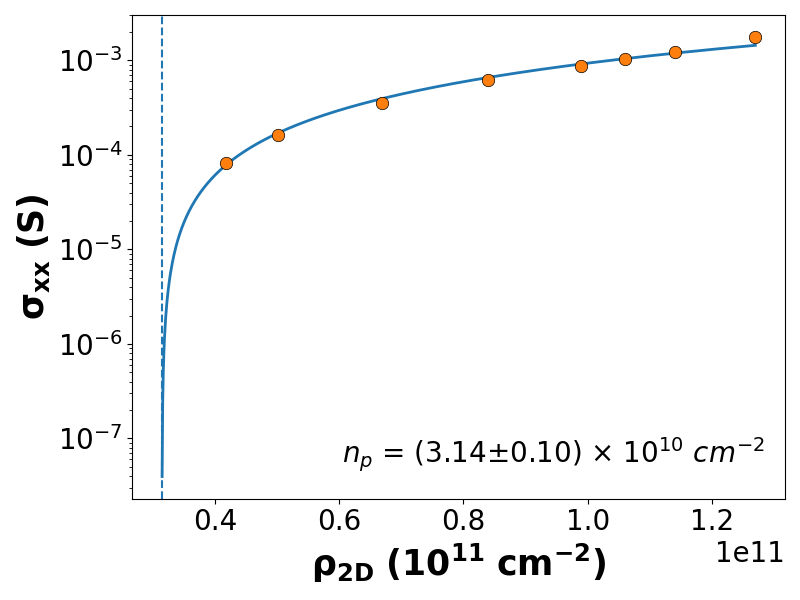}
        \caption{}
        \label{percolation}
    \end{subfigure}

    \caption{\textbf{(a) STEM image} of a FIB sample cut from the Ge/SiGe substrate \textbf{(b) Substrate composition} of FIB sample showing Silicon and Germanium composition of quantum well \textbf{(c) Mobility} vs carrier density. \textbf{(d) Conductivity $\sigma_{xx}$} vs carrier density.}
    \label{fig:material}
\end{figure}

Our reverse-graded Ge/SiGe quantum well is grown on Si wafter via RPCVD, with a target $80\%$ Ge $20\%$ for the SiGe layers, and $100\%$ Ge for the quantum well. STEM image and Ge:Si composition analysis are shown in Fig.~\ref{fig:material}(a) and (b).
We fabricate Hall bar devices on the material and measure mobility with respect to carrier density shown in Fig.~\ref{fig:material}(c). the percolation density was obtained by fitting data in Fig.~\ref{fig:material}(d) to $\sigma_{xx}=A(\rho_{2D}-n_p)^{1.31}$.

\newpage
\section{$T_1$ relaxation}
\label{T1 fitting}
\begin{figure}[htbp]
    \centering
    \begin{subfigure}{0.44\textwidth}
        \centering
        \includegraphics[width=\linewidth]{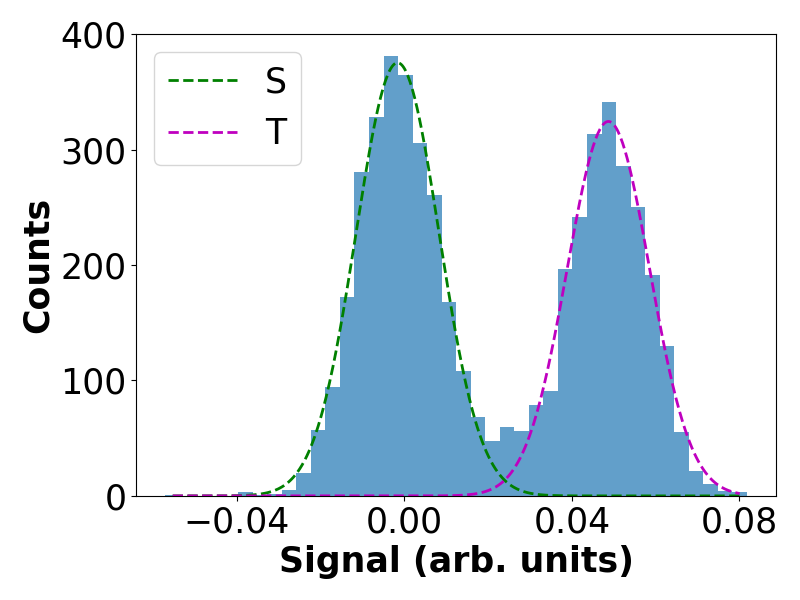}
        \caption{}
        \label{PSB decay 0}
    \end{subfigure}
    \hfill
    \begin{subfigure}{0.45\textwidth}
        \centering
        \includegraphics[width=\linewidth]{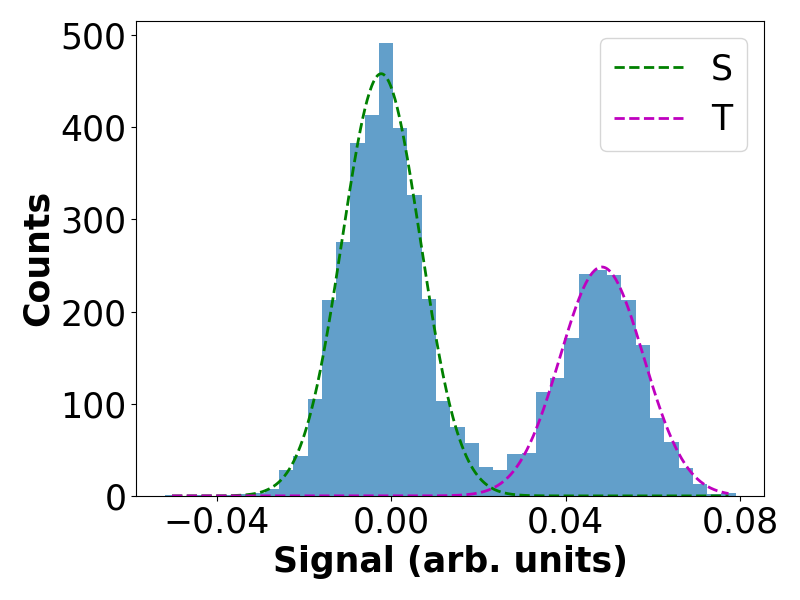}
        \caption{}
        \label{PSB decay 20}
    \end{subfigure}

    \vspace{1em}

    \begin{subfigure}{0.44\textwidth}
        \centering
        \includegraphics[width=\linewidth]{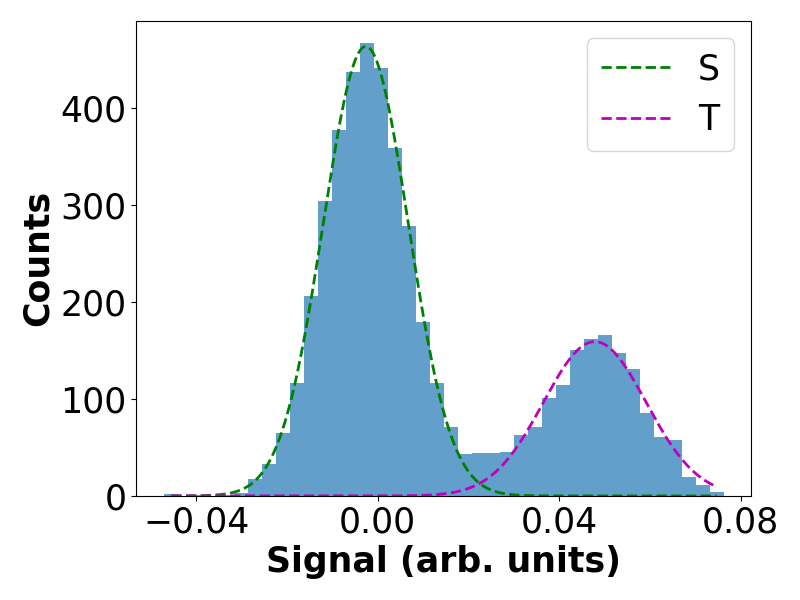}
        \caption{}
        \label{PSB decay 40}
    \end{subfigure}
    \hfill
    \begin{subfigure}{0.45\textwidth}
        \centering
        \includegraphics[width=\linewidth]{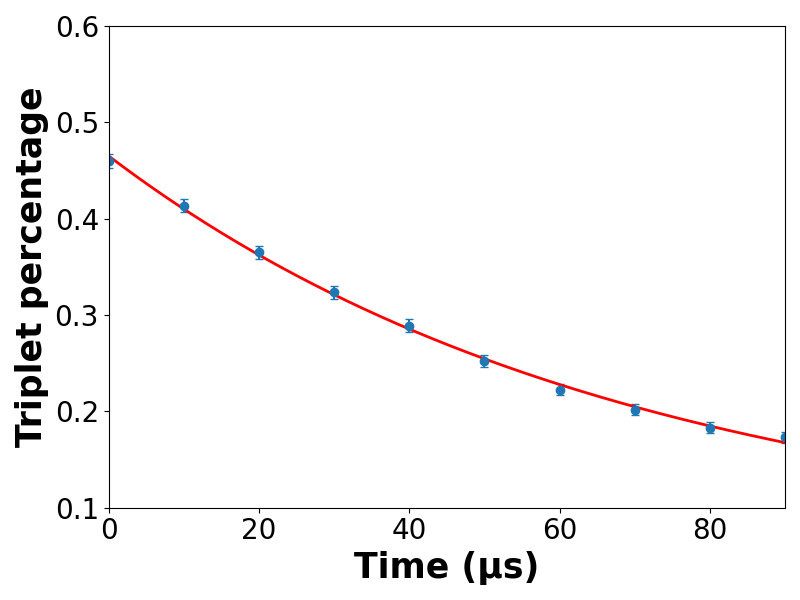}
        \caption{}
        \label{T1}
    \end{subfigure}

    \caption{\textbf{$T_1$ decay fitting} (a),(b),(c) PSB readout signals for 0\,\si{\micro\s}, 20\,\si{\micro\s}, and 40\,\si{\micro\s} wait times respectively. (d) $T_1$ time decay fit}
    \label{fig:T1}
\end{figure}

To obtain $T_1$ decay time of qubits at the readout point, we preform random load experiments where we move to $(0,1)$ charge state, randomly load a spin by moving to $(1,1)$ charge state and then moving to the readout region to preform PSB readout. The triplet states decay naturally to Singlet due to $T_1$ relaxation, and by observing the change in triplet and singlet probabilities for different wait times before readout (Fig~\ref{fig:T1}(a-c)), we can obtain a decay curve with $T_1 = 69(6)$\,\si{\micro\s} (Fig~\ref{fig:T1}(d)).

\newpage
\section{Singlet-triplet qubit}
\label{ST}

 \begin{figure}[htbp]
    \centering
    \begin{subfigure}[b]{0.45\textwidth}
        \centering
        \includegraphics[width=\textwidth]{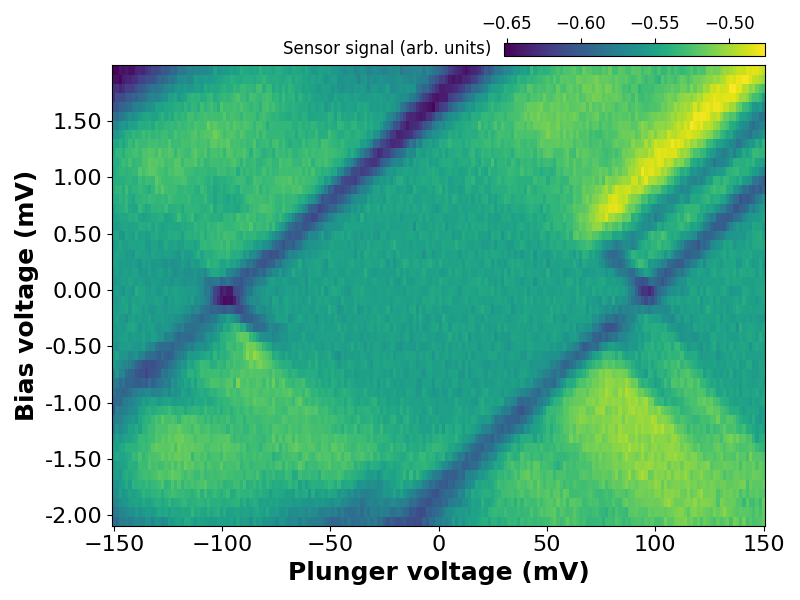}
        \caption{}
        \label{ST coulomb 1}
    \end{subfigure}
    \hfill
    \begin{subfigure}[b]{0.45\textwidth}
        \centering
        \includegraphics[width=\textwidth]{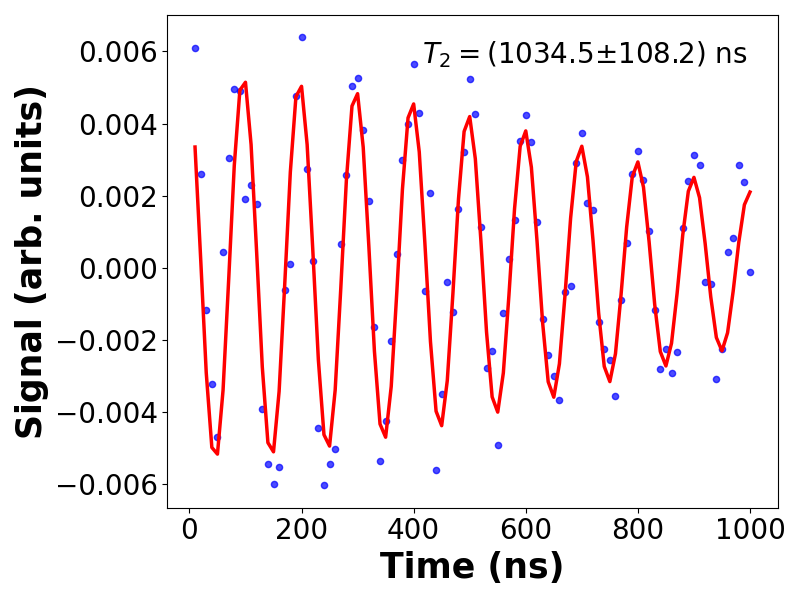}
        \caption{}
        \label{ST}
    \end{subfigure}
    \caption{\textbf{(a) Coulomb diamond} of the sensor. \textbf{(b) ST qubit oscillation} showing coherent ST qubit driving.}
    \label{fig:ST}
\end{figure}

A S-$T_-$ qubit was measured on device 1. Charging energy of the sensor dot is approximately 2\,meV (Fig.~\ref{fig:ST}(a)). Singlet-triplet oscillation measurements are preformed under 50\,mT magnetic field, with the resulting $T^*_2$ measured as $1.0(1)$\,\si{\micro\s} (Fig.~\ref{fig:ST}(b)).
 
\newpage
\section{Charge noise characterization}
\label{charge noise}
\begin{figure}[h]
    \centering
    \begin{subfigure}{0.32\textwidth}
        \centering
        \includegraphics[width=\linewidth]{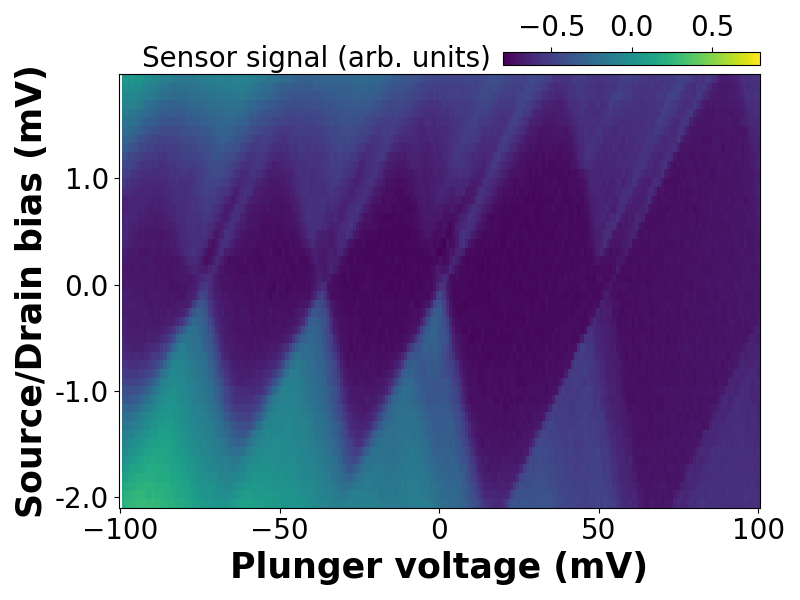}
        \caption{}
        \label{CNCD}
    \end{subfigure}
    \hfill
    \begin{subfigure}{0.32\textwidth}
        \centering
        \includegraphics[width=\linewidth]{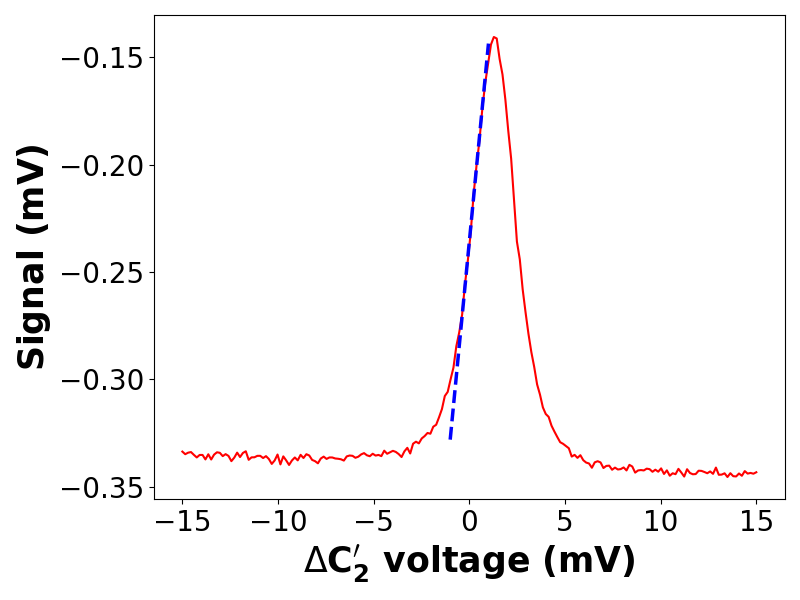}
        \caption{}
        \label{sensor slope}
    \end{subfigure}
    \hfill
    \begin{subfigure}{0.32\textwidth}
        \centering
        \includegraphics[width=\linewidth]{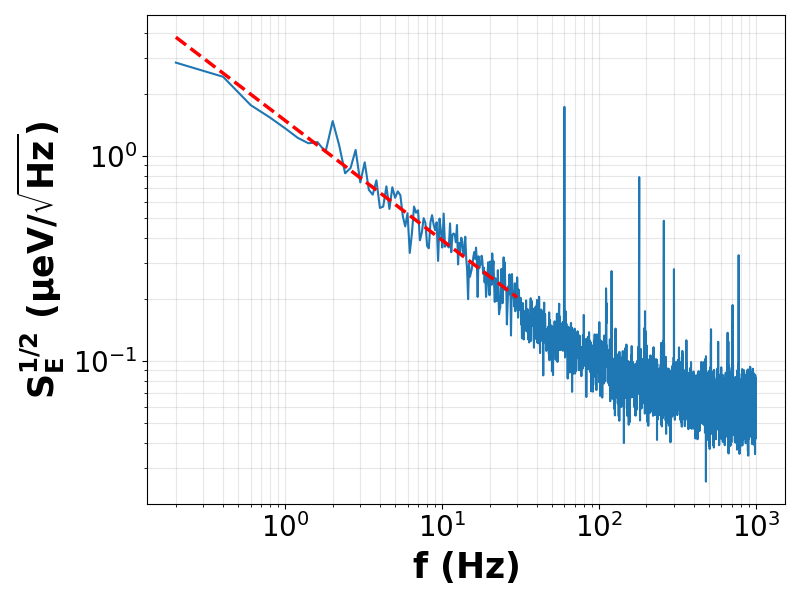}
        \caption{}
        \label{charge noise}
    \end{subfigure}
    \caption{
    \justifying
    \textbf{(a)Coulomb diamond} of the ST qubit, with a extracted leverarm value of $44.6(8)$\,\si{\micro\eV}/mV. \textbf{(b)Sensor signal} of the ST qubit charge sensor fitted at the readout point to obtain signal to plunger voltage ratio of $0.092 (3)$. \textbf{(c)Charge noise} with a $1/f^\gamma$ fit up to 30Hz, resulting in $\gamma = 1.17(3)$ and Charge noise at $1Hz = 1.49(8)$\,\si{\micro\eV}$/\sqrt{Hz}$.} 
    \label{fig:noise}
\end{figure}

We preformed charge noise characterization of the charge sensor dot on device 3. The power spectrum density is obtained from the Fourier transform of a charge sensor signal trace over a period of 10\,s. The Coulomb diamond of the charge sensor (Fig.\ref{fig:noise}(a)) was used to obtain the leverarm while the signal to plunger ratio is obtained from the slope of the Coulomb blockade signal (Fig.\ref{fig:noise}(b)). The resulting charge noise (Fig.\ref{fig:noise}(c)) gives $\gamma$ value of 1.17 and a charge noise of $1.49(8)$ at 1\,Hz.
